\documentclass[nofootinbib]{revtex4}

\usepackage{amssymb}

\newcommand{\be}{\begin{equation}}
\newcommand{\ee}{\end{equation}}
\newcommand{\beqn}{\begin{eqnarray}}
\newcommand{\eeqn}{\end{eqnarray}}
\newcommand{\sltwoR}{$sl(2,{\mathbb R})$}
\newcommand{\slthreeR}{$sl^{(3)}(2,{\mathbb R})$}
\newcommand{\slop}{sl}
\newcommand{\ddx}{\frac{d}{dx}}
\newcommand{\ddxsqr}{\frac{d^2}{dx^2}}
\newcommand{\ra}{\rangle}
\newcommand{\jnull}{j_0}
\newcommand{\jp}{j_+}
\newcommand{\jm}{j_-}
\newcommand{\jpm}{j_{\pm}}
\newcommand{\Jnull}{J_0}
\newcommand{\Jp}{J_+}
\newcommand{\Jm}{J_-}
\newcommand{\Jpm}{J_{\pm}}
\newcommand{\one}{1\!\!1}
\newcommand{\casim}{{\cal C}}

\begin{document}

\author{N. Debergh}
\email{Nathalie.Debergh@ulg.ac.be}
\affiliation{Techniques du son
et de l'image, Montefiore Institute (B28), University of Li\`ege,
B-4000 Li\`ege}
\author{J. Ndimubandi}
\email{jndimubandi@yahoo.fr}
\affiliation{University of Burundi,
Department of Mathematics,
 P.O. Box 2700, Bujumbura (Burundi)}
\author{B. Van den Bossche}
\email{bvandenbossche@ulg.ac.be} \affiliation{Fundamental
Theoretical Physics, Institute of Physics (B5), University of
Li\`ege, B-4000 Li\`ege}

\title{Polynomial deformations of \sltwoR\   in a three-dimensional
invariant subspace of monomials}

\begin{abstract}
New finite-dimensional representations of specific polynomial
deformations of \sltwoR\   are constructed. The corresponding
generators can be, in particular, realized through linear
differential operators preserving a finite-dimensional subspace
of monomials. We concentrate on three-dimensional spaces. 
%and
%point-out new quasi-exactly solvable potentials associated with
%such realizations.
\end{abstract}

\maketitle

\section{Introduction}
\label{Section1}

Quasi-exactly solvable (Q.E.S.) one-dimensional equations have
attracted much attention for the past twenty-five years. For a
review, see f.i. \cite{Ref1}. They are related to quasi-exact
operators, i.e., those which possess a finite-dimensional
invariant subspace coinciding with a representation space of a
finite-dimensional representation of the Lie algebra \sltwoR\  
following Turbiner's statement \cite{Ref2}. Such a feature
implies that quasi-exact operators can be presented in a
block-diagonal matricial form. Consequently, all polynomials of
quasi-exact differentials operators, including some Schr\"odinger
Hamiltonians, are such that a finite number of their eigenvalues
can be analytically determined.

As already stated, the Lie algebra \sltwoR\   seems to be the
cornerstone of Q.E.S. differential operators when one variable
only is under consideration \cite{Ref3}. For completeness, let us
recall that this algebra is generated by $\jnull,\jpm$ such that
\beqn
[\jnull,\jpm]&=&\pm \jpm,\label{Eq1}\\
\left[\jp,\jm\right]&=&2\jnull. \label{Eq2}\eeqn
The $(2j+1)$-dimensional ($j=0,1/2,1,\ldots$) representations are
characterized by
\beqn
\jnull|j,m\ra &=&m|j,m\ra,\label{Eq3}\\
\jpm|j,m\ra&=&\sqrt{(j\mp m)(j\pm m+1)}|j,m\pm 1\ra.\label{Eq4}
\eeqn
The basis ${|j,m\ra,m=-j,\ldots,+j}$ can be explicitly realized
through monomials
\be |j,m\ra=\sqrt{\frac{(2j)!}{(j+m)!(j-m)!}}\  x^{j+m} \label{Eq5}\ee
associated with the quasi-exact operators
\beqn \jnull&=& x\ddx -j,\label{Eq6}\\
\jp&=&-x^2\ddx+2jx,\label{Eq7}\\
\jm&=&\ddx. \label{Eq8}\eeqn

Ad-hoc linear and quadratic combinations of these three operators
lead to the Q.E.S. Schr\"odinger potentials indexed in
\cite{Ref2}.

However, Post and Turbiner asked in \cite{Ref4} for the
possibility of another Lie algebra playing a similar role to the
one of \sltwoR in the context of quasi-exact operators, without
obtaining a satisfactory answer. Even the simplest subspace
\be V_3=\{1,x,x^3\}\label{Eq9} \ee
is not meaningful being preserved by 11 generators (of first-,
second- and third- order) which do not close under the Lie
bracket.

We want to take back the question of Post and Turbiner and prove
that such an algebra does exist, but that it requires to go to
nonlinear structures (up to and included cubic terms). This is
due to the fact that polynomial algebras are characterized by a
wide class of irreductible finite dimensional representations. In
particular, it is known \cite{Ref5} that such representations of
the cubic deformations of \sltwoR, which we name \slthreeR, can
be labeled by two supplementary "quantum numbers" with respect to
the ones of \sltwoR. In Section~\ref{Section2}, we still
enlarge the number of these finite-dimensional representations by
introducing a third supplementary label. We devote a particular
interest to the case of three-dimensional representations having
in mind the space $V_3$. We turn to the corresponding
differential realizations in Section~\ref{Section3}.
% while Section~\ref{Section4} is related to the associated Q.E.S. potentials.
We conclude in Section~\ref{Section5}.

\section{Finite-dimensional representations of \slthreeR}
\label{Section2}

First, we recall that the cubic deformation of \sltwoR, i.e.
\slthreeR, is generated by three operators $\Jnull,\Jpm$ such
that (compare with Eqs.~(\ref{Eq1}) and~(\ref{Eq2}))
\beqn [\Jnull,\Jpm]&=&\pm \Jpm,\label{Eq10}\\
\left[\Jp,\Jm\right]&=&\alpha \Jnull^3+\beta \Jnull^2+\gamma
\Jnull+\delta \label{Eq11}\eeqn
with $\alpha,\beta,\gamma,\delta \in {\mathbb R}$. The Casimir
operator associated with this nonlinear algebra is
\be
\casim=\Jp\Jm+\frac{\alpha}{4}\Jnull^4+\left(\frac{\beta}{3}-\frac{\alpha}{2}\right)\Jnull^3
+\left(\frac{\alpha}{4}-\frac{\beta}{2}+\frac{\gamma}{2}\right)\Jnull^2+
\left(\frac{\beta}{6}-\frac{\gamma}{2}+\delta\right)\Jnull.
\label{Eq12}\ee

Some finite-dimensional representations of \slthreeR\ have
already been pointed out \cite{Ref5}. They are characterized by
\beqn \Jnull|J,M\ra&=&(M/q+c)|J,M\ra,   \label{Eq13}\\
\Jp |J,M\ra&=&F(M)|J,M+q\ra,\label{Eq14}\\
\Jm |J,M\ra&=& G(M)|J,M-q\label{Eq15}\ra,
 \eeqn
with $M=-J,\ldots,+J$ and $J=0,1/2,1,\ldots$. The positive
integer $q$ as well as the real constant $c$ are the
supplementary labels mentioned in the Introduction. They are
constrained according to the dimension of the representations as
well as to the algebra itself. More details on that point and on
the forms of the functions $F(M)$ and $G(M)$ can be found in
\cite{Ref5}.

We now point out a supplementary class of $(2J+1)$-dimensional
representations of \slthreeR. They are such
\beqn \Jnull |J,M\ra&=&\left[a
M^2+(1/q-aq-2aM_1)M+c\right]|J,M\ra,\label{Eq16}\\
\Jp |J,M\ra&=&f(M_1)|J,M+q\ra\delta_{M,M_1},\label{Eq17}\\
\Jm |J,M\ra&=&g(M_1+q)|J,M-q\ra\delta_{M,(M_1+q)}.
\label{Eq18}\eeqn
While $M=-J,\ldots,+J$, the integer $M_1$ is a fixed value in the
range $M_1\in[-J,J]$.  For instance, Eq.~(\ref{Eq17}) means that
$\Jp|J,M\ne M_1\ra=0$ and $\Jp|J,M=M_1\ra=f(M_1)|J,M_1+q\ra$, i.e.,
$\Jp|J,M\ra$ is vanishing for the $2J$ values $M\ne M_1$ and is
nonvanishing only for $M=M_1$. A similar conclusion holds
for~(\ref{Eq18}). Relations~(\ref{Eq16}) to~(\ref{Eq18}) have
been found in order to fulfill~(\ref{Eq10}) as the reader can be
convinced starting from $\Jnull|J,M\ra=(aM^2+bM+c)|J,M\ra$ and
acting both sides of~(\ref{Eq10}) on $|J,M\ra$, thus giving the
constant $b$. Then, acting with $|J,M_1+q\ra,|J,M_1\ra$ and
$|J,\tilde{M}\ra$ on both sides of~(\ref{Eq11}), where $\tilde{M}$
is neither equal to $M_1+q$ nor $M_1$,  it turns out that
Eq.~(\ref{Eq11}) is satisfied if
\beqn
f(M_1)g(M_1+q)&=&\alpha\left(c+M_1/q+1-aqM_1-aM_1^2\right)^3+\beta\left(c+M_1/q+1-aqM_1-aM_1^2\right)^2
\nonumber\\
&&\hspace{0cm}\mbox{}+
\gamma\left(c+M_1/q+1-aqM_1-aM_1^2\right)+\delta,\label{Eq19}\\
f(M_1)g(M_1+q)&=&-\alpha\left(c+M_1/q-aqM_1-aM_1^2\right)^3-\beta\left(c+M_1/q-aqM_1-aM_1^2\right)^2\nonumber\\
&&\hspace{0cm}\mbox{}-
\gamma\left(c+M_1/q-aqM_1-aM_1^2\right)-\delta,\label{Eq20}\\
0&=&\alpha\left[c+\left(1/q-aq-2aM_1\right)\tilde{M}
+a\tilde{M}^2\right]^3
+\beta\left[c+\left(1/q-aq-2aM_1\right)\tilde{M}+a\tilde{M}^2\right]^2
\nonumber\\
&&\hspace{0cm}\mbox{}+\gamma\left[c+\left(1/q-aq-2aM_1\right)\tilde{M}
+a\tilde{M}^2\right]^3+\delta,\label{Eq21}
%\quad\forall M\ne M_1, M\ne M_1+q, 
\eeqn
respectively. The constraint~(\ref{Eq19}) or~(\ref{Eq20}) 
fixes the action of the
ladder operator $\Jpm$ on the basis while
Eqs.~(\ref{Eq20}) or~(\ref{Eq19}) and~(\ref{Eq21}) are $2J$ constraints on the
labels of the representations as well as the algebra itself\footnote{
Eq.~(\ref{Eq21}) is an ensemble of $(2J+1)-2$ constraints. The elimination
of $f(M_1)g(M_1+q)$ between Eqs.~(\ref{Eq19}) and~(\ref{Eq20}) gives a supplementary condition, hence the mentioned $2J$ constraints.}.

\section{Three-dimensional differential realization of \slthreeR}
\label{Section3}

We now turn to a three-dimensional space by fixing $J=1$.
According to Eqs.~(\ref{Eq17})--(\ref{Eq18}), we have three
possibilities %
\beqn q&=&1,M_1=-1,\label{Eq22}\\
q&=&1,M_1=0,\label{Eq23}\\
q&=&2,M_1=-1.\label{Eq24} \eeqn
The fact that we have only these three possibilities is because
$M=-1,0,1$, hence $M_1$ can only be fixed to $M_1=-1, M_1=0$ or
$M_1=1$. Since $\Jp$ is a raising operator and we span a space of
three states, $q$ can only be chosen to be $q=1$ or $q=2$. This
immediately forbids the choice $M_1=1$ because the state
$|J,M_1+q\ra=|J,1+q\ra$ is not in the basis. With this reasoning
with a ladder operator, it is easy to see that for $q=1$ we can
only have $M_1=-1$ and $M_1=0$, while with $q=2$, only the choice
$M_1=-1$ is allowed.

In this Section we are
interested in the three-dimensional differential realizations of
\slthreeR\ and more specifically on those acting in a space of
monomials. As already discussed in \cite{Ref1}, two states of the
space can be fixed to $1$ and $x$ without loosing generality. The
justification lies on the freedom of performing a change of
variable and a change of eigenfunction, two modifications leaving
the eigenvalues unchanged. For simplification, we shall
concentrate on $V_3$.

Due to the relation~(\ref{Eq10}), $\Jnull$ can be realized with a
first order operator only. We then have
\be \Jnull=\frac{1}{p}x\ddx+C \label{Eq25}\ee
while we can make the following identifications
\be |1,-1\ra\rightarrow 1,\quad |1,0\ra\rightarrow x,\quad
|1,1\ra\rightarrow x^3. \label{Eq26}\ee
Together with Eq.~(\ref{Eq16}), we obtain
\be a=\frac{1}{2p},\quad C=c-\frac{1}{p}\label{Eq27}\ee
with
\be p=\frac{1}{2}q^2+q\left(M_1+3/2\right). \label{Eq28}\ee

So far, each of the three possibilities~(\ref{Eq22})--(\ref{Eq24}) 
is characterized 
by one label $c$ (see~(\ref{Eq16}), with $a$ now fixed 
according to~(\ref{Eq27})). The parameters $\alpha,\beta,\gamma,\delta$ being
not yet fixed, the algebra is thus still
arbitrary at that stage. In other words, every \slthreeR\ admits
three-dimensional representations of the
type~(\ref{Eq16})--(\ref{Eq18}). 

Let us discuss the three cases~(\ref{Eq22})--(\ref{Eq24}) in
details.

\subsection{The case $\left(q=1,M_1=-1\right)$}

Such a case implies $p=1$, thus $a=1/2$. Eq.~(\ref{Eq19}) is then
\be f(-1)g(0)=\alpha c^3+\beta c^2+\gamma c+\delta,
\label{Eq29}\ee
while Eqs.~(\ref{Eq20})--(\ref{Eq21}) give rise to the two
constraints
\beqn 2\alpha
c^3+(2\beta-3\alpha)c^2+(2\gamma+3\alpha-2\beta)c+2\delta-\gamma+\beta-\alpha&=&0,\label{Eq30}\\
\alpha(2+c)^3+\beta(2+c)^2+\gamma(2+c)+\delta&=&0.\label{Eq31}
\eeqn
Two quantities are thus fixed. We choose them to be the label $c$
and the parameter $\delta$:
\be c=-\frac{7}{10}-\frac{\gamma}{2\beta},\quad
\delta=\frac{\gamma^2}{4\beta}-\frac{169}{100}\beta,\quad
\mbox{if } \alpha=0, \label{Eq32} \ee
and \beqn c&=&
-\frac{7}{10}-\frac{\beta}{3\alpha}\pm\frac{1}{30\alpha}\sqrt{-579\alpha^2+100\beta^2-300\alpha\gamma},
\nonumber\\
\delta&=&\frac{39}{125}\alpha-\frac{2}{27}\frac{\beta^3}{\alpha^2}
+\frac{\beta\gamma}{3\alpha}\pm\left(\frac{\beta^2}{135\alpha^2}-\frac{\gamma}{45\alpha}-\frac{166}{1125}
\right)\sqrt{-579\alpha^2+100\beta^2-300\alpha\gamma}, \quad
\mbox{if } \alpha\ne0.\label{Eq33} \eeqn
The case $(\alpha,\beta)=(0,0)$ has been excluded because leading
to $\gamma=\delta=0$ and thus to a triviality.

The diagonal operator~(\ref{Eq25}) with $c$ arbitrary just obeys~(\ref{Eq16}).
Moreover, if we request it to realize the diagonal
operator $\Jnull$ of \slthreeR, this fixes $c$ (and $\delta$)
according to~(\ref{Eq32}) or~(\ref{Eq33}). We still have to analyse the
ladder operators. Relations~(\ref{Eq17})--(\ref{Eq18})
are satisfied on the basis~(\ref{Eq26}) with
\beqn
\Jp&=&f(-1)\left(\frac{1}{3}x^3\ddxsqr-x^2\ddx+x\right),\label{Eq34}\\
\Jm&=&g(0)\left(-\frac{1}{2}x\ddxsqr+\ddx\right).\label{Eq35}
\eeqn
For the moment, the coefficients are arbitrary. However, if the
relation~(\ref{Eq29}) is taken into account,
these realizations are such that the algebra \slthreeR\ with
$\delta$ fixed by~(\ref{Eq32}) or~(\ref{Eq33}) is actually
generated by the operators~(\ref{Eq34})--(\ref{Eq35}) on the
space $V_3$. However, nothing says that it is still the case in
an intrinsic way. In fact, one can easily be convinced that it is
effectively true if
\be \gamma=-\frac{31}{16}\alpha+\frac{\beta^2}{3\alpha}
\label{Eq36} \ee
and $\alpha<0$ $(\alpha>0)$ for the upper (lower) sign of
Eq.~(\ref{Eq33}).

We conclude that the operators~(\ref{Eq34}) and~(\ref{Eq35}) with
\be f(-1)g(0)=\frac{3}{2}\alpha \label{Eq37}\ee
and the operator
\be \Jnull=x\ddx-\frac{7}{4}-\frac{\beta}{3\alpha}\label{Eq38} \ee
not only preserve $V_3$, but also generate the \slthreeR\ algebra
characterized by Eqs.~(\ref{Eq10}) and
\be
[\Jp,\Jm]=\alpha\Jnull^3+\beta\Jnull^2+\left(\frac{\beta^2}{3\alpha}
-\frac{31}{16}\alpha\right)\Jnull
+\frac{15}{32}\alpha-\frac{31}{48}\beta+\frac{\beta^3}{27\alpha^2}.\label{Eq39}
\ee
The Casimir operator~(\ref{Eq12}) reduces to a multiple of the
identity operator:
\be
\casim=\left(\frac{315}{1024}\alpha-\frac{23}{48}\beta+\frac{23}{288}\frac{\beta^2}{\alpha}
+\frac{\beta^3}{54\alpha^2}-\frac{\beta^4}{324\alpha^3}\right)\one. \ee

We also notice that the three-dimensional representation of the
\slthreeR\ algebra in Eqs.~(\ref{Eq10}) and~(\ref{Eq39}) and
subtended by the operators~(\ref{Eq34})-(\ref{Eq35}) and~(\ref{Eq38})
is reducible. It actually reduces to a direct sum of two irreducible
representations of this algebra: One one-dimensional representation
(corresponding to $J=0$) and one two-dimensional representation 
(corresponding to $J=1/2$). We summarize this situation by
\be
\left(J=1,c=-\frac{3}{4}-\frac{\beta}{3\alpha}\right)=
\left(J=0,c=\frac{5}{4}-\frac{\beta}{3\alpha}\right)\oplus
\left(J=\frac{1}{2},c=-\frac{a}{4}-\frac{5}{4}-\frac{\beta}{3\alpha}\right).
\ee
It corresponds to the fact that within the space $V_3$, the element 
$x^3$ is separated from the two others.

\subsection{The case $\left(q=1,M_1=0\right)$}

This case leads to $p=2$ and $a=1/4$. Eq.~(\ref{Eq19}) is
\be f(0)g(1)=\alpha (c+1)^3+\beta (c+1)^2+\gamma (c+1)+\delta,
\label{Eq40}\ee
and the corresponding two constraints are
\beqn 2\alpha
c^3+(2\beta+3\alpha)c^2+(2\gamma+3\alpha+2\beta)c+2\delta+\gamma+\beta+\alpha&=&0,\label{Eq41}\\
\alpha(c-1/2)^3+\beta(c-1/2)^2+\gamma(c-1/2)+\delta&=&0.\label{Eq42}
\eeqn
The label $c$ and the parameter $\delta$ are fixed according to
\be c=-\frac{1}{8}-\frac{\gamma}{2\beta},\quad
\delta=\frac{\gamma^2}{4\beta}-\frac{25}{64}\beta,\quad \mbox{if }
\alpha=0, \label{Eq43} \ee
and \beqn c&=&
-\frac{1}{8}-\frac{\beta}{3\alpha}\pm\frac{1}{24\alpha}\sqrt{-111\alpha^2+64\beta^2-192\alpha\gamma},
\nonumber\\
\delta&=&-\frac{15}{128}\alpha-\frac{2}{27}\frac{\beta^3}{\alpha^2}
+\frac{\beta\gamma}{3\alpha}\pm\left(\frac{\beta^2}{108\alpha^2}-\frac{\gamma}{36\alpha}-\frac{47}{1152}
\right)\sqrt{-111\alpha^2+64\beta^2-192\alpha\gamma}, \quad
\mbox{if } \alpha\ne0.\label{Eq44} \eeqn

Relations~(\ref{Eq17})--(\ref{Eq18}) 
%with~(\ref{Eq40}) 
on the basis~(\ref{Eq26})
lead to
\beqn
\Jp&=&f(0)\left(-\frac{1}{2}x^4\ddxsqr+x^3\ddx\right),\label{Eq45}\\
\Jm&=&g(1)\left(\frac{1}{6}\ddxsqr\right).\label{Eq46} \eeqn
These operators, together with $\Jnull$ of Eqs.(\ref{Eq25}), (\ref{Eq27}) 
and~(\ref{Eq28}), realize \slthreeR\ on the space $V_3$ 
if Eqs.~(\ref{Eq40})  and~(\ref{Eq44}) or~(\ref{Eq45}) are taken into account.
Moreover,
they realize \slthreeR\ independently of the space iff
\be \gamma=-\frac{5}{8}\alpha+\frac{\beta^2}{3\alpha} \label{Eq47}
\ee
with $\alpha>0$ $(\alpha<0)$ for the upper (lower) sign of
Eq.~(\ref{Eq44}).

Summarizing, the operators~(\ref{Eq45}) and~(\ref{Eq46}) with
\be f(0)g(1)=\frac{3}{16}\alpha \label{Eq48}\ee
together with the operator
\be
\Jnull=\frac{1}{2}x\ddx-\frac{1}{2}-\frac{\beta}{3\alpha}\label{Eq49}
\ee
not only stabilize $V_3$ but also  generate  \slthreeR\ with
\be
[\Jp,\Jm]=\alpha\Jnull^3+\beta\Jnull^2+\left(\frac{\beta^2}{3\alpha}-\frac{5}{8}\alpha\right)\Jnull
-\frac{3}{16}\alpha-\frac{5}{24}\beta+\frac{\beta^3}{27\alpha^2}.\label{Eq50}
\ee
The Casimir operator~(\ref{Eq12}) reduces again to the identity
operator up to a prefactor:
\be 
\casim=\left(-\frac{\beta}{24}+\frac{\beta^2}{144\alpha}
+\frac{\beta^3}{54\alpha^2}-\frac{\beta^4}{324\alpha^3}\right)\one. 
\ee

We point out once again the reducibility of the three-dimensional
 representation subtended by these developments.
More precisely we have
\be
\left(J=1,c=-\frac{\beta}{3\alpha}\right)=
\left(J=0,c=-\frac{1}{2}-\frac{\beta}{3\alpha}\right)\oplus
\left(J=\frac{1}{2},c=-\frac{a}{4}+\frac{1}{2}-\frac{\beta}{3\alpha}\right),
\ee
and the element $1$ is separated from the two others in $V_3$.

\subsection{The case $\left(q=2,M_1=-1\right)$}

We have $p=3$ and $a=1/6$. The constraint on the ladder operator
is
\be f(-1)g(1)=\alpha (c+2/3)^3+\beta (c+2/3)^2+\gamma
(c+2/3)+\delta, \label{Eq51}\ee
while those on $c$ and $\delta$ read
\beqn 2\alpha
c^3+(2\beta+\alpha)c^2+(2\gamma+5\alpha/3+2\beta/3)c+2\delta+\gamma/3
+5\beta/9+7\alpha/27&=&0,\label{Eq52}\\
\alpha c^3+\beta c^2+\gamma c+\delta&=&0.\label{Eq53}
\eeqn
The two last equations imply
\be c=-\frac{5}{6}-\frac{\gamma}{2\beta},\quad
\delta=\frac{\gamma^2}{4\beta}-\frac{25}{36}\beta,\quad \mbox{if }
\alpha=0, \label{Eq54} \ee
and \beqn c&=&
-\frac{5}{6}-\frac{\beta}{3\alpha}\pm\frac{\sqrt{3}}{18\alpha}\sqrt{47\alpha^2+12\beta^2-36\alpha\gamma},
\nonumber\\
\delta&=&\frac{5}{3}\alpha-\frac{2}{27}\frac{\beta^3}{\alpha^2}
+\frac{\beta\gamma}{3\alpha}\pm\frac{1}{\sqrt{3}}\left(\frac{\beta^2}{27\alpha^2}
-\frac{\gamma}{9\alpha}-\frac{34}{81}
\right)\sqrt{47\alpha^2+12\beta^2-36\alpha\gamma}, \quad \mbox{if
} \alpha\ne0.\label{Eq55} \eeqn

From
Eqs.~(\ref{Eq17})--(\ref{Eq18}) and the basis~(\ref{Eq26}), we obtain
\beqn
\Jp&=&f(-1)\left(\frac{1}{3}x^5\ddxsqr-x^4\ddx+x^3\right),\label{Eq56}\\
\Jm&=&g(0)\left(\frac{1}{6x}\ddxsqr\right).\label{Eq57} \eeqn
These operators, together with $\Jnull$ of Eqs.(\ref{Eq25}), (\ref{Eq27}) 
and~(\ref{Eq28}), realize \slthreeR\ on the space $V_3$ if Eqs.~(\ref{Eq51}) 
and~(\ref{Eq54}) or~(\ref{Eq55}) are taken into account.
They satisfy the commutation relations of \slthreeR\ independently of the space
for
\be \gamma=-\frac{55}{144}\alpha+\frac{\beta^2}{3\alpha}
\label{Eq58} \ee
and $\alpha>0$ $(\alpha<0)$ for the upper (lower) sign of
Eq.~(\ref{Eq55}).

In conclusion, the operators~(\ref{Eq56}) and~(\ref{Eq57}) with
\be f(-1)g(1)=-\frac{\alpha}{18} \label{Eq59}\ee
and the operator
\be
\Jnull=\frac{1}{3}x\ddx-\frac{5}{12}-\frac{\beta}{3\alpha}\label{Eq60}
\ee
not only preserve $V_3$ but also generate  \slthreeR\ with
\be
[\Jp,\Jm]=\alpha\Jnull^3+\beta\Jnull^2+\left(\frac{\beta^2}{3\alpha}-\frac{55}{144}\alpha\right)\Jnull
-\frac{\alpha}{32}-\frac{55}{432}\beta+\frac{\beta^3}{27\alpha^2}.\label{Eq61}
\ee
The associated Casimir operator still reduces to the identity
operator up to a prefactor:
\be 
\casim=\left(-\frac{23}{432}\beta-\frac{17}{2592}\frac{\beta^2}{\alpha}
+\frac{\beta^3}{54\alpha^2}-\frac{\beta^4}{324\alpha^3}-\frac{1045}{82944}\alpha\right)\one.
\ee

Like in the two previous cases,  the three-dimensional
 representation is reducible following
\be
\left(J=1,c=-\frac{1}{6}-\frac{\beta}{3\alpha}\right)=
\left(J=0,c=-\frac{1}{12}-\frac{\beta}{3\alpha}\right)\oplus
\left(J=\frac{1}{2},c=-\frac{a}{4}+\frac{1}{12}-\frac{\beta}{3\alpha}\right),
\ee
and corresponding in $V_3$ to the separation of $x$ from the two other 
elements.

%\section{Quasi-exact potentials associated with \slthreeR}
%\label{Section4}

\section{Conclusions}
\label{Section5}

We have answered the Post and Turbiner question asking for an
algebra generated by the differential operators preserving $V_3$.
This algebra is not a Lie structure but a polynomial deformation
of it. The representation of these nonlinear algebras are however
available and can be used as the ones of the Lie algebra are. In
particular, it is thinkable to go to higher-dimensional
representations of \slthreeR\ and ask for a differential
realization associated with a space of specific interest. Anyway,
it is the main advantage of having recognized an algebraic
structure behind these quasi exact operators.

We have also, using this algebraic approach, recovered the
Post-Turbiner operators~(\ref{Eq34})--(\ref{Eq35}),
(\ref{Eq45})--(\ref{Eq46}), (\ref{Eq56}) while the
operator~(\ref{Eq57}) is new. They are in fact separated in three
pairs generating a specific \slthreeR\ together with an ad-hoc
diagonal operator. We however note that the complete set of these
six operators does not close under the Lie bracket even if a
nonlinear deformation is considered. The explanation is simple:
Acting on $V_3$ these six operators together with two adequate
diagonal ones generate $\slop(3,{\mathbb R})$. This latest Lie
algebra being realized with differential operators in terms of two
variables, it is not conceivable to hope a closure of the
nonlinear deformation of $\slop(3,{\mathbb R})$ with one variable
only. The extension of the developments contained in this letter 
to a space of arbitrary finite dimensions is left for future work.

\end{document}